\documentclass[aps,prb,twocolumn,groupedaddress]{revtex4}
\usepackage{graphicx}
\usepackage{epsfig}
\usepackage{float}

\newcommand{\be}{\begin{equation}}
\newcommand{\ee}{\end{equation}}
\begin{document}
\title{On Heavy Carbon Doping of MgB$_{2}$}
\author{Deepa Kasinathan, K.-W. Lee, and W.E. Pickett}

\affiliation{Department of Physics, University of California, 
          Davis CA 95616}

\begin{abstract} 
Heavy carbon doping of MgB$_2$ is studied by first principles electronic 
structure studies of two types, 
an ordered supercell (Mg(B$_{1-x}$C$_{x}$)$_{2}$, 
$x$=0.0833) and
also the coherent potential approximation method that incorporates
effects of B-C disorder.
For the ordered model, the twofold degenerate $\sigma$-bands that are the 
basis of the high temperature superconductivity 
are split by 60 meV ({\it i.e.} 7 meV/\% C) and the $\sigma$ 
Fermi cylinders contain 0.070 holes/cell, compared to 0.11 for MgB$_2$. 
A virtual crystal treatment tends to overestimate the rate at which $\sigma$
holes are filled by substitutional carbon.
The coherent potential approximation (CPA) calculations give the same
rate of band filling as the supercell method.  The occupied local density
of states of C is almost identical to that of B in the upper 2 eV of the
valence bands, but in the range -8 eV to -2 eV, C has a considerably
larger density of states.  The calculations indicate that the $\sigma$
Fermi surface cylinders pinch off at the zone center only above the
maximum C concentration $x \approx$ 0.10.  These results indicate that
Mg(B$_{1-x}$C$_{x}$)$_{2}$ as well as Mg$_{1-x}$Al$_x$B$_2$ is a good
system in which to study the evolution of the unusual electron-phonon
coupling character and strength as the crucial $\sigma$ hole states are
filled.

\end{abstract}
\maketitle
\today

\section{Introduction}
Discovery of superconductivity at T$_c \approx$ 40 K in MgB$_2$ has 
enlivened not only interest in new classes of superconductors with high
T$_c$ and novel two-gap behavior, but also pursuit of new materials for
applications that require high critical current densities and high
critical fields (H$_{c2}$).  In highly resistive films (due to unreacted 
components and/or oxygen and carbon impurities) the perpendicular and
parallel critical fields have reached\cite{gurevich} $H_{c2}^{\perp}
\approx$ 34 T, $H_{c2}^{\parallel}\approx$ 49 T, making MgB$_2$ 
a real possibility for a high field conductor. 
Such application require defects or grain boundaries to pin the 
vortices whose motion would otherwise lead to energy dissipation and joule
heating.  Such defects also affect the underlying electronic structure and
pairing interaction, at least in some parts of the sample, and often
decrease T$_c$.  The simplest defect to understand is the substitutional 
impurity, which provides a means of varying the intrinsic properties in a
continuous and controllable manner and may enhance pinning mechanisms.  
Although readily synthesizable as a
stoichiometric, rather clean compound, the MgB$_2$ lattice resists most
attempts to alloy, both on the Mg and B sites, at more than the very
dilute level.

The two established exceptions are Al substitution for Mg, and C substitution
for B, both of which lead to a rapid decrease in T$_c$.  Published
reports on the behavior vary considerably with synthesis method and sample
treatment, but some of the general behavior seems to be established.  Al
substitution should be the simpler one, since the very strongly bonded 
honeycomb structure B layers remain intact, and simple rigid band filling
of the hole states at first appears to be a likely possibility.  
Up to $y = 0.10$ in Mg$_{1-y}$Al$_y$B$_2$ that appears to be the case,
with T$_c$ decreasing smoothly.  Beyond the concentration
$y = 0.10$, however, there are signs of two-phase behavior 
(two $c$ lattice parameters in diffraction studies) arise.\cite{cava1}
Around $y = 0.25$ the system reverts to a single phase very low and
finally vanishing T$_c$.  A study of the trends in electron-phonon coupling
strength using first principles calculations in the virtual crystal 
approximation\cite{profeta} reported a sharp change of behavior at $y$=
0.25.  A thorough study of the energetics of Al substitution revealed a
strong tendency for superstructure formation\cite{barabash}
at $y$=0.25 (and $y$=0.75).
Still, the observed  
onset of two-phase behavior already at $y$=0.10 is unexplained, and 
nonstoichiometry\cite{grin} and microscopic defects must be kept in mind.

Unlike the Al alloying case where there has been at least rough consensus on
changes of properties, reports of the change in T$_{c}$ and structure 
with addition
of carbon on MgB$_{2}$ have varied widely. These 
differences seem in some cases to reflect real differences in materials
due to the various methods of synthesis and heat treatment, and the 
samples prepared were highly mixed phase, thereby 
creating additional uncertainty.  The very similar xray scattering strength
of the B and C atoms has rendered standard xray diffraction ineffective in
determining the C content of a sample.  
Reports early on suggested C miscibilities in Mg(B$_{1-x}$C$_x$)$_2$ as 
small\cite{maurin} as
$x$=0.02 to as large\cite{bharathi} as $x$=0.3.  Recent studies
of several groups suggest that $x$=0.2 can be achieved while larger
concentrations are questionable\cite{schmidt,adveev, ribiero} although
reports of larger C concentrations under high pressure growth 
persist.\cite{masui}

With some of the materials questions coming under control, the 
observation trends
are raising some serious questions.  The decrease in T$_c$ follows $dT_c/dx 
\sim 1$ K/\% C,[\onlinecite{wilke}] with several reports of 
T$_c$ $\approx$ 21-23 K at 
$x$=0.2.[\onlinecite{ribiero, adveev,schmidt}]  However, with carbon adding one
electron and the $\sigma$ bands of MgB$_2$ holding only 
$\sim$0.11 holes per unit cell,\cite{jan}
one might guess that the $\sigma$ bands would be filled (or nearly so, as the
$\pi$ bands can also accept carriers) and superconductivity would have
vanished.  Not only is T$_c$ still robust at $x$=0.2, analysis of tunneling 
spectra indicates two-band superconductivity is 
retained,\cite{holanova,samuely} whereas it might also
seem likely that disorder scattering should have averaged out the gap to
a single value.  The upper critical field H$_{c2}$ initially {\it increases}
strongly with C content as T$_c$ is depressed, reaching a 
maximum\cite{wilke,gurevich,puzniak}
of $\sim$33-35 T.  This critical field is substantially less than that
for more disordered films (see above).

Recent studies seem to agree
that C can be introduced substitutionally for B  
up to 10\% replacement 
$x$=0.10.\cite{avdeev,schmidt}
Tunneling spectroscopies indicate that Mg(B$_{1-x}$C$_x$)$_2$ remains a
two gap superconductor\cite{schmidt, holanova} ({\it i.e.} the gap 
anisotropy is not washed out by scattering) for 10$\pm$2\% C substitution 
for B.  This question of how the anisotropy gets washed out has attracted
much interest.  Similarly, the mechanisms underlying the high critical
fields are not understood.  These questions are not the focus 
of the present study, although the knowledge gained from first principles
calculations will be useful in the resolution of these topics.

A few first principles calculations have been 
done for C substitution of B in MgB$_2$.   Pseudopotential calculations
including relaxation in a 27 unit cell supercell\cite{yanfa} confirmed that 
substitutional C is energetically favorable to interstitial C, and 
that the C-B bondlength is 5-6\% shorter than the bulk B-B bondlength. 
A projector-augmented-wave calculation\cite{erwin} reported a smaller
relaxation, but the difference may be due to constraints related
to supercell size.
A coherent potential approximation study of the disordered alloy using
a Korringa-Kohn-Rostoker multiple scattering method in the atomic sphere
approximation\cite{ppsingh} revealed relatively small effects of disorder
in the range 0$\le x \le 0.3$.
The main effect of C substitution was reported to be the raising of the Fermi 
energy due to the additional carriers, but the filling of the crucial
$\sigma$ hole band was not quantified. 

In this paper, we present a more detailed analysis of the changes in
electronic structure, and the effective $\sigma$ band doping, using
both periodic supercells and disordered alloy calculations. 
Our study lays the groundwork for another issue in the 
superconductivity of MgB$_2$ alloys that has received no direct attention from
experimentalists: the effect of decreasing holes on the strength and
character of electron-phonon coupling.  Theoretical studies\cite{libc1,2D,kong}
have predicted the following remarkable changes in doped MgB$_2$: as the
number of holes decreases (hence the Fermi wavevector $k_F$
of the cylindrical $\sigma$ Fermi
surfaces decreases), (1) the substantial downward renormalization of the
E$_{2g}$ bond-stretching modes with Q$\le 2k_F$ does not change, (2)
the coupling strengths of these already strongly coupled modes {\it increases},
and (3) the total coupling strength $\lambda$, and hence T$_c$, remains
unchanged.  This scenario assumes two-dimensionality of the $\sigma$ bands
which is only approximately true, and neglects the change of electron-phonon
matrix elements which should be reasonable.  This argument also neglects
the fact that the applicability of conventional electron-phonon 
(Midgal-Eliashberg) theory, already somewhat suspect\cite{boeri,wep}
in MgB$_2$, definitely becomes inapplicable as the Fermi energy decreases 
and approaches more closely the bond-stretching mode energy (65 meV).
Item (2) definitely portends unusual dynamics related to the B-B
stretch modes.
We hope the current paper stimulates experimental investigation into 
these questions. 
   
\section{Calculational Methods}
Two methods of assessing the effects of substitutional C in MgB$_2$ 
have been used in the work reported here.
Our ordered impurity calculations have been performed
using the full-potential linearized augmented
plane wave code Wien2k\cite {wien}, applying the
generalized gradient approximation\cite{gga} to the exchange-correlation
potential. The basis set is reduced by using the APW+lo
method\cite{apw}, retaining the accuracy of the LAPW
method. RK$_{\it max}$ was set to 7.00 which is a 
high quality basis for the $s$-$p$ electron systems. The Brillouin zone 
integration for the supercell was carried out using 432  k points in the
irreducible part of the zone. We have used the experimental
results of Avdeev {\it et al.}\cite {avdeev} 
for the lattice parameters of 
the 8.33\% doped supercell and the 10\% doped virtual
crystal. 

For the disordered (randomly substituted) alloy calculations,
we have used the full-potential nonorthogonal local-orbital minimum-basis
scheme (FPLO)\cite{klaus1} applied in the coherent 
potential approximation (CPA).\cite{klaus2}
In the calculations, 1152 ($45\times 45\times 10$) 
irreducible k points, 
and valence orbitals $\it{2s2p3s3p3d}$ for Mg and
$\it{2s2p3d}$ for B and C, were used.
The implementation of the CPA in FPLO relies on the Blackman-Esterling-Berk
theory\cite{BEB} that includes random off-diagonal matrix elements in the
local orbital representation.

\section{Ordered M{\lowercase{g}}(B$_{1-x}$C$_{x}$)$_2$, $x$ = 0.0833}
\begin{figure}[tb]
\psfig{figure=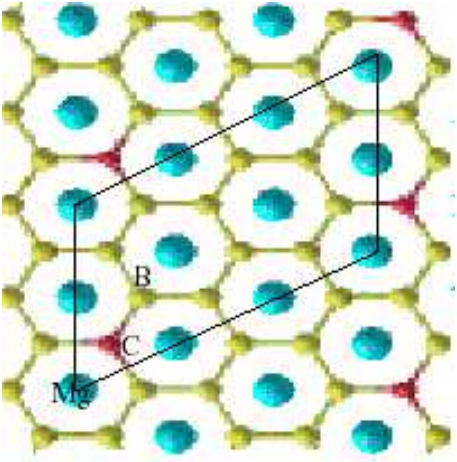,width=6cm}
\caption{(Color online) Top view of the B-C layer of the
Mg$_{6}$B$_{11}$C supercell.
Large (blue) circles denote
Mg atoms, small light (yellow) circles denote boron,
and small dark (red) circles denote carbon. The supercell boundary is
outlined.}
\label{supercell}
\end{figure}

Smaller concentrations of ordered C substitution for B 
require larger supercells.
Our choice of supercell was guided by the desire for a small enough C
concentration to be relevant to address experimental data, yet not so small
as to make it difficult to distinguish the effect of C addition or to
make the calculations unreasonably tedious.  Since interlayer hopping is
not an issue we will address, the supercell will involve enlargement
of only a single 
B layer.  The shape of supercell 
should have as small as possible aspect ratio in order to maximize the
separation of C atoms.
The compromise we chose was the 8.33\%
doped supercell (Mg$_{6}$B$_{11}$C $\rightarrow$ 
Mg(B$_{1-x}$C$_{x}$)$_{2}$, x = 1/12 = 0.0833).
Figure \ref{supercell} shows the top view of the B$_{11}$C layer in 
the 2$\times$3 supercell. 
Each carbon atom has three first, 
five second, and three third nearest 
neighbor boron atoms. 

\begin{table}[tb]
\caption{Lattice parameter values used in the virtual crystal and supercell
calculations.}
\begin{center}
\begin{tabular}{|c|c|c|c|}\hline
Lattice & undoped & 10\% doping & 8.33\% doping \\
Parameter & MgB$_{2}$ & MgB$_{1.8}$C$_{0.2}$\cite {avdeev}
 & Mg$_{6}$B$_{11}$C \\
 & ($\AA$) & ($\AA$) & ($\AA$) \\ \hline \hline
$a$ & 3.083 & 3.053 & 3$\times$3.053=9.159 \\ \hline
$b$ & 3.083 & 3.053 & 2$\times$3.053=6.106 \\ \hline
$c$ & 3.521 & 3.525 & 3.525 \\ \hline
\end{tabular}
\end{center}
\end{table}

\subsection{Band structure}
There are two effects of carbon substitution: a change in the average
potential in the B-C layer, and the breaking of symmetry by C replacement
of B in the supercell.  In the ordered supercell the potential difference
is kept explicit and symmetry breaking will be clear.  In Sec. VI 
CPA will be used to probe these effects in a
different manner.  Another effect, B relaxation around the C impurities,
has been addressed previously to some extent, but it will not be studied 
here since the effect cannot be included within the CPA.

\begin{figure}[tb]
\psfig{figure=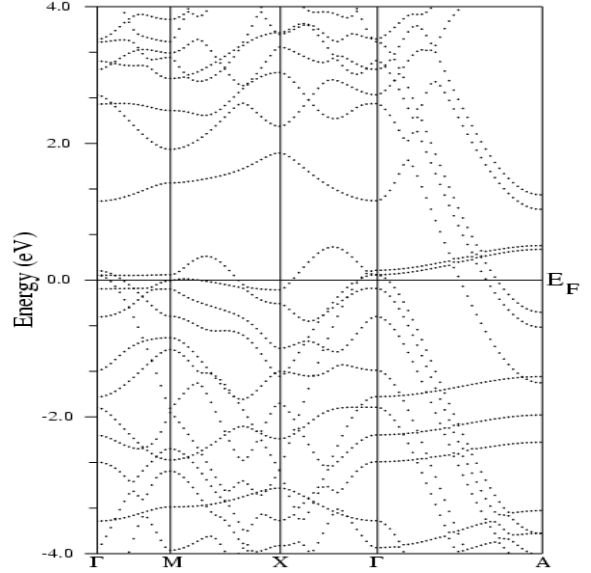,height=8cm,width=3in}
\caption{Band structure of Mg$_{6}$B$_{11}$C in the region of the Fermi level. 
The twofold degenerate $\sigma$ band
in the undoped system along the $\Gamma$-A
is split by 60 meV when 1/12 of B is replaced by C.}
\label{bands}
\end{figure}

The band structure within 4 eV of the Fermi
level (E$_{F}$) is shown in Fig \ref{bands}. 
The primary band in the understanding of MgB$_2$
superconductivity is the (twofold degenerate) 
$\sigma$ band along the $\Gamma$-A direction.
The results of supercell band-folding can be distinguished especially for
the $\Gamma$-A direction, where (for example) the $\sigma$ band is 
replicated at $\Gamma$ at -1.8 eV, -2.3 eV, -2.6 eV, and even further 
lower energies.  

\begin{center}
\begin{figure}[tb]
\psfig{figure=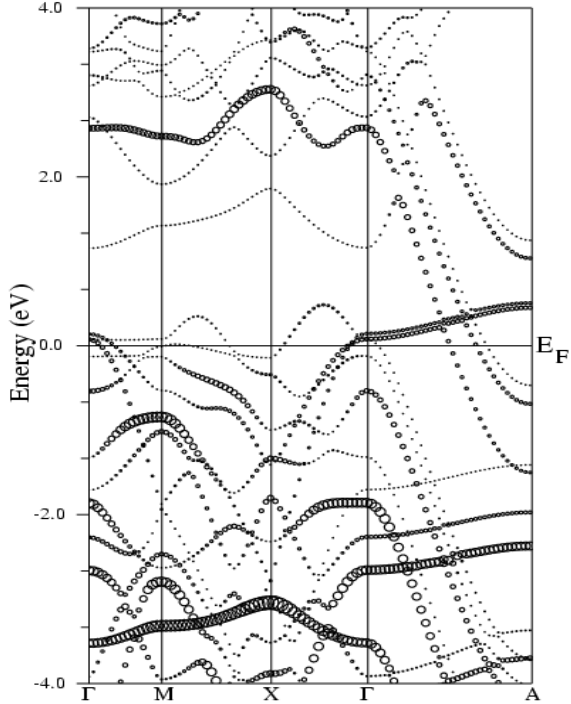,height=10cm,width=3in}
\caption{Bands of Mg$_6$B$_{11}$C near the Fermi level, with 
symbol size proportional to the C $2p$ character.  The $\sigma$ bands,
split by 60 meV (hardly visible in this figure) lie just above E$_F$ at 
$\Gamma$ and disperse upward toward the A point.}
\label{fatbands}
\end{figure}
\end{center}

One of the most readily apparent effects of C doping is the splitting of
the $\sigma$ bands along $\Gamma$-A, by 60 meV.  It is helpful in 
understanding this splitting to obtain the atomic characters of each of
the split bands, which will in any case involve only the B and C $p\sigma$
($p_x, p_y$) states.
The carbon $p$ contribution is emphasized by enlarged symbols (``fatbands'')
in Fig. \ref{fatbands}.
The lower of the two bands has stronger C content 
(although this is not easy to distinguish
in Fig. \ref{fatbands}).  In addition, it has more first and second B 
neighbor character than does the upper band.  The upper band contains more
third neighbor B character, this being the B site farthest from the C atom
and representative of the bulk material.
The interpretation of this splitting is that lower band has been pulled
down due to the stronger potential of the C atom compared to that of B. 
The 60 meV splitting for $x$=1/12 provides (assuming a linear effect in
this concentration range) an energy scale for $\sigma$-band broadening
$\gamma_{\circ}\approx $ 7 meV/\% C content.

\subsection{Density of States}
\begin{center}
\begin{figure}[tb]
\psfig{figure=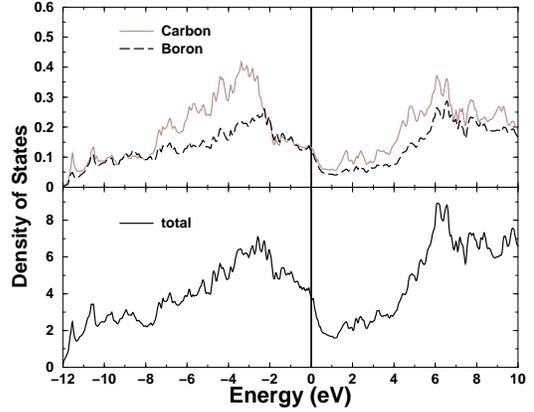,angle=-00,width=7cm}
\caption{Density of states, total (bottom) and decomposed into B and C
contributions on a per-atom basis, for the ordered $x$=1/12 model. The
primary B -- C difference is that the C DOS is lower in the -2 eV to E$_F$
region, but higher in the lower region -8 eV to -2 eV. }
\label{DOS}
\end{figure}
\end{center}

Figure \ref{DOS} shows the total and atom-projected 
density of states (DOS) of the
8.33\% doped system. 
The C and B DOS are similar at and above E$_F$, but the C DOS is somewhat
lower in the interval within 2 eV below E$_F$. 
The C DOS is larger than that of B in the
-8 eV to -3 eV region.
Based on charge within
spheres of 1.65$\AA$ radius we calculated a charge transfer of about
0.095e$^{-}$ to the carbon from the three first
nearest neighbor boron atoms, with other B sites showing negligible
change in charge.  This charge transfer also reflects the stronger
potential and larger electronegativity of C, and provides substitutional
C with definite anionic character.

\section{10\% doping}
\begin{figure}[tb]
\psfig{figure=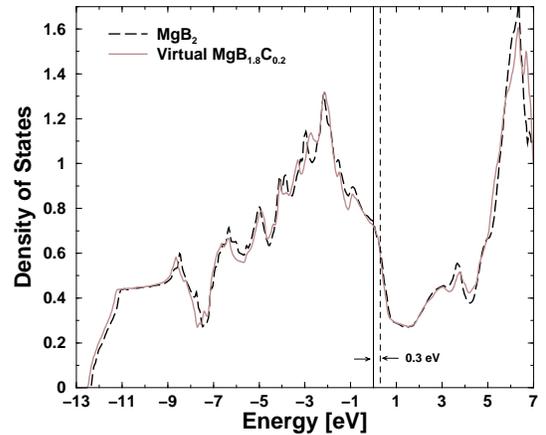,width=7cm}
\caption{Density of States of Mg(B$_{0.9}$C$_{0.1}$)$_2$. The
Fermi level moves up by 0.3eV due to doping}
\label{vcaDOS}
\end{figure}

\begin{figure}[tb]
\psfig{figure=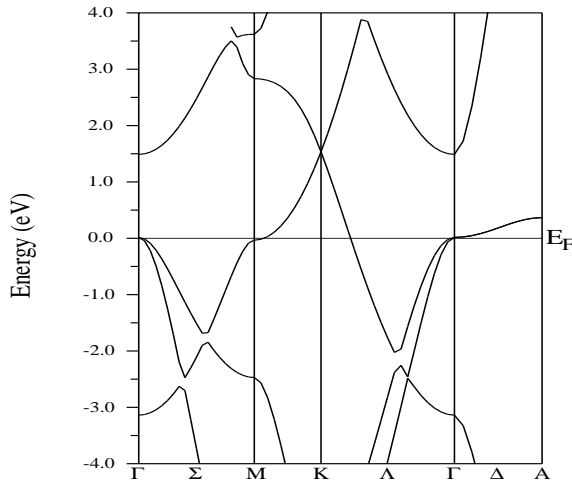,height=8cm,width=3.8in}
\caption{Band structure of Mg(B$_{1-x}$C$_{x})_2, x=0.10$, in the virtual crystal
approximation.  The $\sigma$ bands at $\Gamma$ lie precisely at E$_F$,
so the Fermi cylinder radii at $\Gamma$ vanish, corresponding to a 
topological transition. }
\label{vcabands}
\end{figure}

We have also performed a virtual crystal calculation for the 10\%
doped $x$=0.10 system Mg(B$_{0.9}$C$_{0.1}$)$_2$. The DOS and band structure
are shown in Figs. \ref{vcaDOS} and \ref{vcabands}. By aligning the highest 
peaks in the DOS at -2 eV and +6 eV (they can be aligned simultaneously), we
establish that the differences are (1) an increase in the occupied bandwidth
(at the MgB$_2$ band filling) from 12.4 eV to 12.5 eV ($\sim$1\%), and (2)
the raising of the Fermi level
by 0.3 eV to accommodate the extra electrons.  
Otherwise there is very little difference in
the two densities of states.  This raising of E$_F$ will be compared to
the supercell result in the next section. 
At this band filling, the VCA gives the $\sigma$ band edge at $\Gamma$ 
precisely at E$_F$ as shown in Fig. \ref{vcabands}.  
At the point A it is still about 0.3 eV above E$_F$, meaning that
both cylindrical Fermi surfaces have shrunk to a point at their ``waists.''

\section{$\sigma$ hole concentration}
It is accepted that superconductivity in MgB$_{2}$ arises from
hole-doping of the $\sigma$ bonding hole states due to the intrinsic
chemistry\cite{jan} of MgB$_2$.  The observed decrease in electron-phonon
coupling strength, and hence T$_c$ upon C addition, makes the
change in hole concentration one of the major points of interest.
We have calculated
the hole concentration of both the 8.33\% supercell system
and the 10\% virtual system from the volume enclosed by the Fermi
surface.
Since the radii of the cylindrical Fermi surfaces are only
slightly different along the $\Gamma \rightarrow$ M and the
$\Gamma \rightarrow$ K directions, we calculated the average
basal area by considering the value along the $\Gamma \rightarrow$
M direction only. Also, we assumed a sinusoidal dependence of
the Fermi surface along the $\hat c$ axis, allowing analytic evaluation
of the Fermi surface volumes {\it i.e.} the hole concentrations. The results
are presented in Table II. In the process of calculating
the number of holes, electronic structure reveals that 
both the $\sigma$ band Fermi surfaces are still intact,
consistent with the two-band superconductivity with
substantial T$_{c}$ as seen in experiments.\cite{schmidt}
Also, we interpolated the number of holes 
for the 10\% doped system (virtual crystal) to calculate
the value for the 8.33\% doped system. Supercell calculation
gave 0.070 holes/cell while the extrapolation gave only 0.057 holes/cell.
This indicates that the virtual crystal approximation
is not very reliable for substitutional carbon, and that C cannot
be thought of ``boron + an electron.''

\begin{table}[t]
\caption{Calculated number of $\sigma$ holes for the various
C concentrations discussed in the text.}
\begin{center}
\begin{tabular}{|c|c|} \hline
 & No. of holes \\
 & (holes/cell) \\ \hline \hline
MgB$_{2}$ & 0.11 \\ \hline
8.33\% doping & \\ 
Mg$_{6}$B$_{11}$C supercell & 0.070 \\ \hline
10\% doping & \\
MgB$_{1.8}$C$_{0.2}$ virtual crystal & 0.0463 \\ \hline
8.33\% doping & \\
MgB$_{1.833}$C$_{0.167}$ extrapolation & 0.057 \\ \hline
\end{tabular}
\end{center}
\end{table}

\section{Coherent Potential Approximation Results}
\begin{center}
\begin{figure}[itb]
\psfig{figure=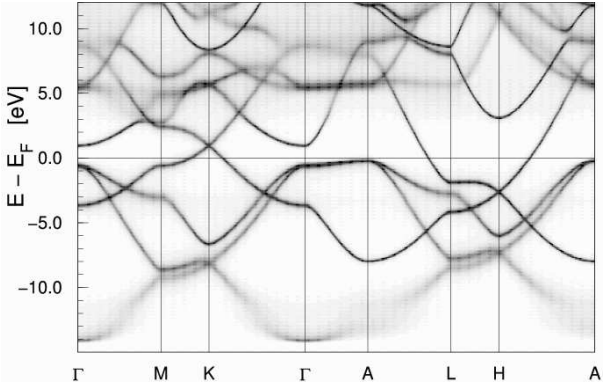,width=8cm}
\caption{CPA spectral density in the full valence-conduction band region,
for $x$=0.20, plotted as a broadened band structure.  Disorder
broadening is largest below -8 eV in the valence bands and above
4 eV in the conduction bands.}
\label{fullCPA}
\end{figure}
\end{center}

For further comparison and to assess the effects of disorder, we have
performed CPA studies of this system.
We have carried out CPA calculations (1) at $x$=0.0001 to provide a ``perfect 
crystal'' reference for evaluation of the various algorithms in the CPA
code, (2) at $x$ = 0.0833 for most direct comparison to the $x$=1/12 (ordered)
supercell calculations described in the previous sections, and (3) at $x$=
0.10 and 0.20 as
representative of the system toward the upper range of achievable C 
substitution.  The spectral function over the whole energy range of
interest, plotted as a ``smeared'' band structure, is shown for the 
$x$=0.20 case in Fig. \ref{fullCPA}.  The primary points of interest are
the filling of the $\sigma$ band hole states, and the broadening (and 
potentially splitting) of bands. 

The full energy region is presented in Fig. \ref{fullCPA} 
for $x$=0.20 (where broadening is more easily seen) to illustrate
the strong wavevector (k) and energy (E) dependence of the 
broadening of the spectral
function.  The largest disorder occurs in the $2s$ region in the lower
valence band, where the C $2s$ state is noticeably lower in energy
leading to increased smearing.
The other region of large disorder is from +3 eV upward, but some
states remain comparatively sharp.  The flat band at 5 eV along 
$\Gamma$-M and A-L, with strong $\pi$ character, is the lowest conduction
band that is strongly affected by the disorder.
The bands around the Fermi level, whether $\sigma$ or $\pi$, are
among those {\it less affected} by the chemical disorder.
As expected from the virtual crystal results for $x$=0.10 from the previous
section, the $\sigma$ band holes are completely filled at $x$=0.20,
as revealed by the flat band along the $\Gamma$-A line lying entirely 
below E$_F$.  
The band filling behavior versus $x$ is quantified below.

In the valence (occupied) bands,
disorder broadening is large in the lower $s$-band region -8 eV to -14 eV
below E$_F$, and somewhat less so
where the (primarily $\sigma$) band in the -3 to -4 eV range gets flat
around the zone edge M and L points.  Similar broadening does not occur in
the same band at the (more distant from $\Gamma$) 
zone edge K and H points, where the
band lies at -6 eV and is much less flat.  The top of the $\sigma$ 
bonding bands along $\Gamma$-A are comparatively sharp; the width is
quantified below.

In the conduction bands disorder broadening becomes more prevalent.
The antibonding $\sigma^*$ bands (flat along $\Gamma$-A at +6 eV) are
much broader than their bonding counterparts (below E$_F$), and this
band along $\Gamma$-M and A-L  with Mg $3s$ character
becomes exceedingly diffuse.  The $\pi$
bands show strong k-dependence of the broadening, beginning at 2-3 eV 
around the M point and becoming wider in the 6-10 eV range 
(and above, not shown in the figure).  

\begin{figure}[tb]
\begin{center}
\psfig{figure=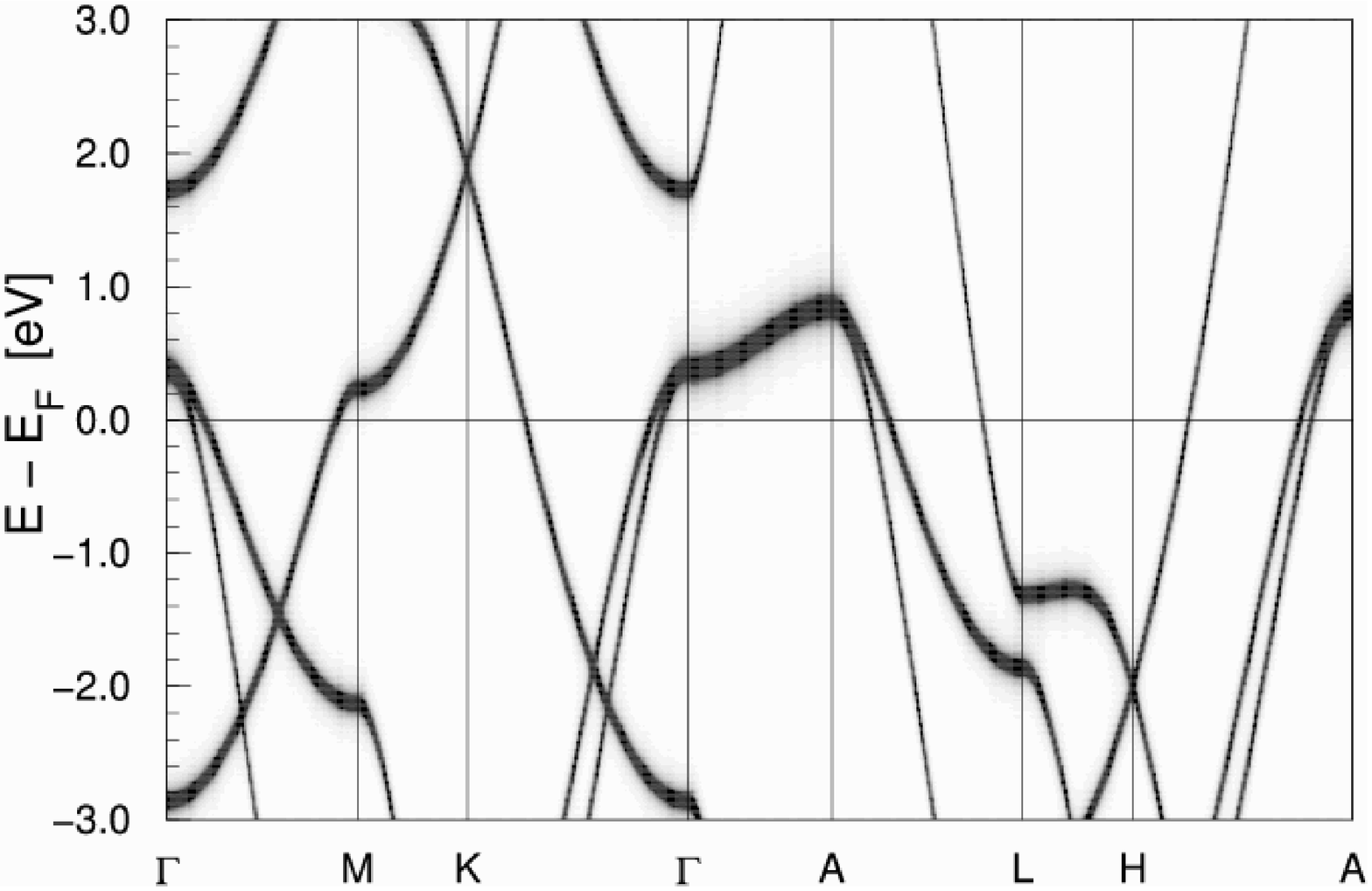,width=7cm}
\psfig{figure=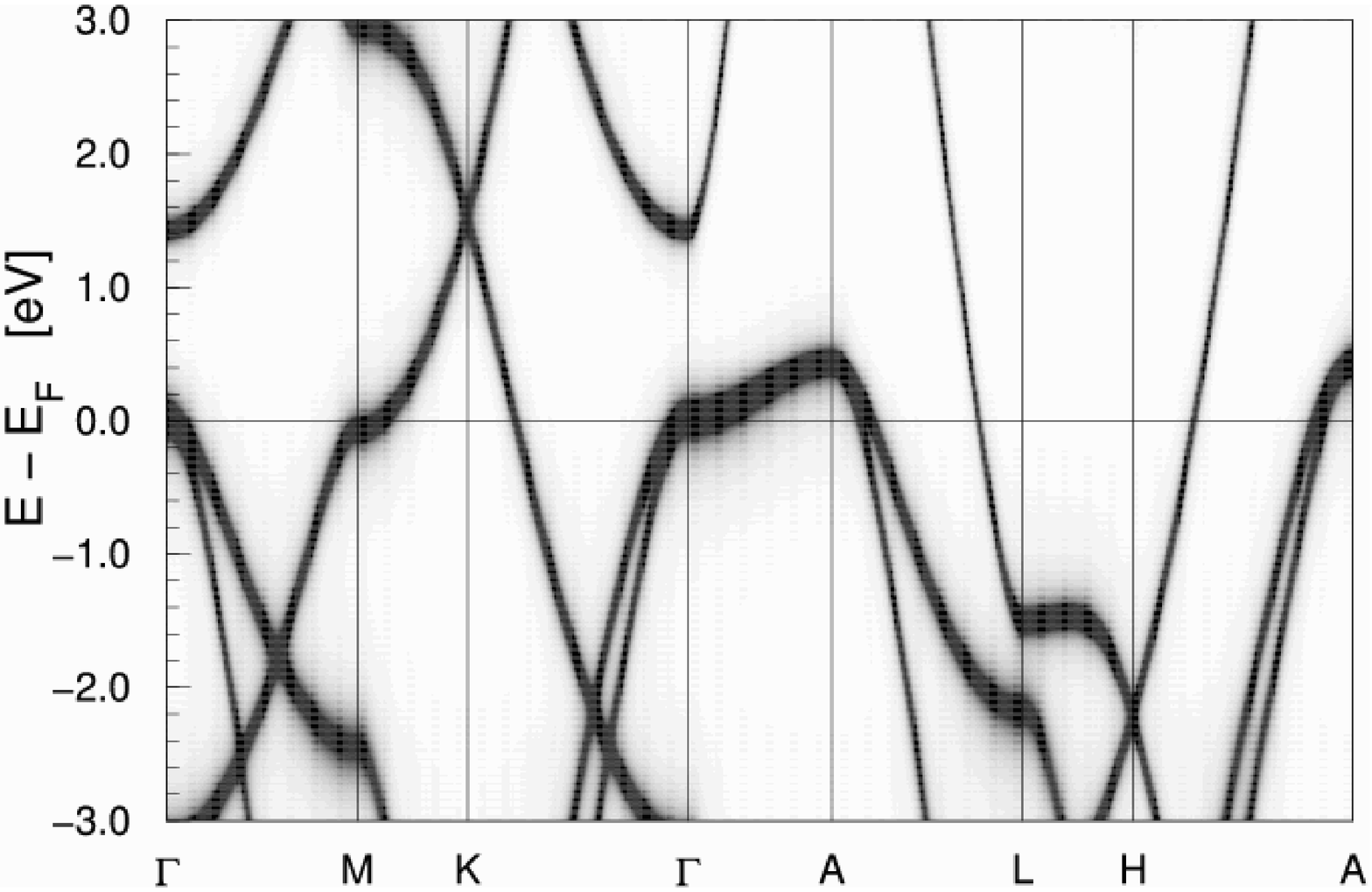,width=7cm}
\psfig{figure=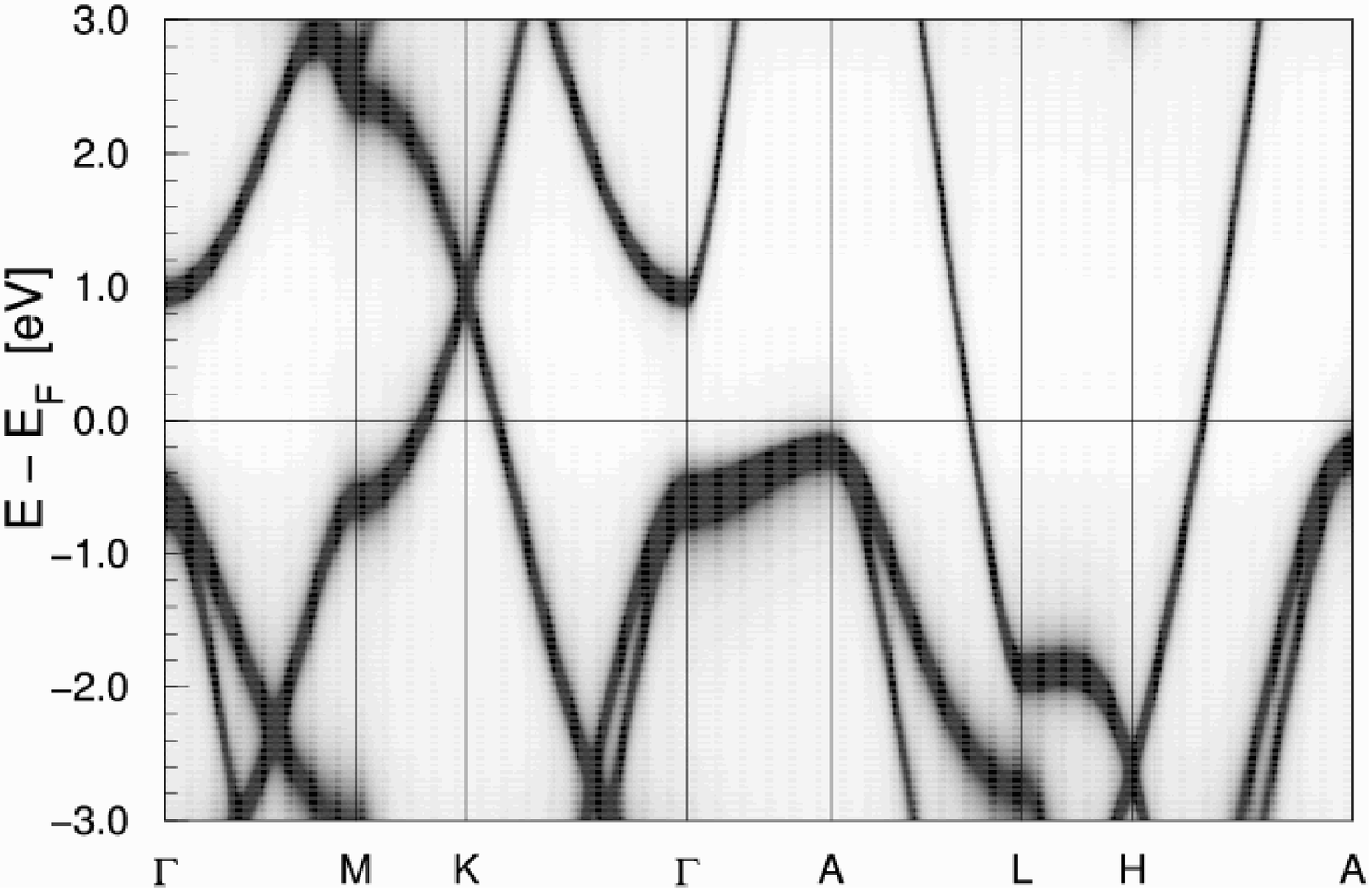,,width=7cm}
\caption{CPA spectral density for $x$=0.0001 (top panel),
$x$=0.10 (middle) and $x$=0.20 (bottom panel).  The main features are
(i) the ``rising'' Fermi level (with respect to the $\sigma$ band along
$\Gamma$-A, say), and (ii) the increase in the disorder broadening of
the bands.  Since the numerical algorithms cannot reproduce the 
$\delta$-function bands for $x\rightarrow 0$, the $x$=0.0001 case
is included as a reference for the algorithmic contribution to the width.}
\label{smallCPA}
\end{center}
\end{figure}

{\it Carbon concentration dependence.} The $x$ dependence of the broadening
can be seen in Fig. \ref{smallCPA} for $x$=0.0001 (the CPA equivalent of
$x$=0 MgB$_2$), for $x$=0.10, and for $x$=0.20 (the latter may not be
experimentally accessible).  Since in each case E$_F$ is set to zero, band
filling appears as downward shifts of the bands, by roughly $\delta$E$_F$
= -0.4 eV for $x$=0.10 and $\delta$E$_F$ = -1.0 eV for $x$=0.20.  For
$x$=0.0833 (not shown) the $\sigma$ bands lie at the same 
position (with respect to E$_F$) as for the ordered $x$=1/12 case shown
in Fig. \ref{fatbands}.  Hence the degree of band filling in the CPA
results is the same as for the ordered $x$=1/12 case that was analyzed
in the previous section.  Although we will not dwell on it, it can be
noticed that the band shift is not entirely rigid: the two valence bands
at the L point split apart as well as broaden with increasing C concentration.

The other important aspect of the CPA bands of Fig. \ref{smallCPA} is the
broadening that increases with carbon concentration.  Within the resolution
of widths that we are able to extract, the widths of the $\sigma$ bands
can be taken to be proportional to $x$,
but differ considerably between $\Gamma$ and A (being about 50\% wider at
$\Gamma$).  The full width at half maximum of the spectral density of the
$\sigma$ bands, averaged between $\Gamma$ and A, is about 
$\gamma \approx $0.21 eV (to perhaps 10\% accuracy).  
This width corresponds to a width in wavevector
given by $\gamma = v_F \delta k$, and therefore a mean free path of
$\ell_F = 2\pi v_F/\gamma$.  The 60 meV splitting of the $\sigma$ bands 
(see Sec. III.A)
for the $x=1/12$ ordered supercell implies that the 0.21 eV mean width can be 
interpreted as 0.05 eV from the B/C on-site energy difference, and thus
0.16 eV from disorder itself. 

Due to the anisotropy of the Fermi velocity $v_F$, the mean free path 
may vary considerably over the Fermi surface at $x$=0.10.  At $\Gamma$ the
cylinder radius has shrunk to a point, and the very small $z$ component
of $v_F$ (vanishing at $\Gamma$) suggests a very small mean free path
in the $z$ direction of the order of the layer spacing $c$ for $x$=0.10.

\section{Discussion and Summary}
In the band structure of the 8.33\% doped supercell
system, the hole $\sigma$-band Fermi surfaces along the 
$\Gamma \rightarrow$ A direction are still present, and this degree
of band filling is reproduced by the CPA.  Even for $x$=0.10 the 
virtual crystal picture
(which overestimates the rate of band filling by C) leaves $\sigma$ Fermi
surfaces, just beginning to be pinched off.  The CPA calculations 
(Fig. \ref{smallCPA}) show the $\sigma$ band holes begin to disappear
rapidly for $x > 0.10$.  Qualitatively this filling is consistent
with most experimental reports.

The questions, and experimental probes, should now be focused on the
fact that, as the $\sigma$ band fills, there will be very strong
deviation from `business as usual' in the coupled electron-phonon
system.  The strength of 
coupling of bond-stretching modes with Q$< 2k_F$ continues to increase, and 
conventional Migdal-Eliashberg theory ceases to apply.  The dynamics of
these ultra-strongly coupled modes is unexplored, with their peculiar
character being signaled by the divergence of their linewidth (at least
within Migdal-Eliashberg theory).  The limiting behavior should not
revert to the widely studied polaron limit, however, as there remain the
background $\pi$ electrons, which are weakly coupled to vibrations but
provide full metallic conductivity and screening to the system.
The present study reveals that C substitution for B provides a similarly
favorable system to Al substitution for Mg for studying this evolution.
Recent reports indicate that Sc substitution (Mg$_{1-x}$Sc$_x$B$_2$) may
also provide\cite{bianconi} another such system for study.

\section{Acknowledgments}
We acknowledge communication on the topics of this paper with
P. C. Canfield, L. C. Cooley, K. Koepernik, I. I. Mazin, and D. J. Singh.
This work was supported by National Science
Foundation Grant DMR-0421810.


\begin{thebibliography}{32}
\expandafter\ifx\csname natexlab\endcsname\relax\def\natexlab#1{#1}\fi
\expandafter\ifx\csname bibnamefont\endcsname\relax
  \def\bibnamefont#1{#1}\fi
\expandafter\ifx\csname bibfnamefont\endcsname\relax
  \def\bibfnamefont#1{#1}\fi
\expandafter\ifx\csname citenamefont\endcsname\relax
  \def\citenamefont#1{#1}\fi
\expandafter\ifx\csname url\endcsname\relax
  \def\url#1{\texttt{#1}}\fi
\expandafter\ifx\csname urlprefix\endcsname\relax\def\urlprefix{URL }\fi
\providecommand{\bibinfo}[2]{#2}
\providecommand{\eprint}[2][]{\url{#2}}

\bibitem[{\citenamefont{Gurevich et~al.}(2004)\citenamefont{Gurevich, Patnaik,
  Braccini, Kim, Mielke, Song, Cooley, Bu, Kim, Choi et~al.}}]{gurevich}
\bibinfo{author}{\bibfnamefont{A.}~\bibnamefont{Gurevich}},
  \bibinfo{author}{\bibfnamefont{S.}~\bibnamefont{Patnaik}},
  \bibinfo{author}{\bibfnamefont{V.}~\bibnamefont{Braccini}},
  \bibinfo{author}{\bibfnamefont{K.~H.} \bibnamefont{Kim}},
  \bibinfo{author}{\bibfnamefont{C.}~\bibnamefont{Mielke}},
  \bibinfo{author}{\bibfnamefont{X.}~\bibnamefont{Song}},
  \bibinfo{author}{\bibfnamefont{L.~D.} \bibnamefont{Cooley}},
  \bibinfo{author}{\bibfnamefont{S.~D.} \bibnamefont{Bu}},
  \bibinfo{author}{\bibfnamefont{D.~M.} \bibnamefont{Kim}},
  \bibinfo{author}{\bibfnamefont{J.~H.} \bibnamefont{Choi}},
  \bibnamefont{et~al.}, \bibinfo{journal}{Supercond. Sci. Technol.}
  \textbf{\bibinfo{volume}{17}}, \bibinfo{pages}{278} (\bibinfo{year}{2004}).

\bibitem[{\citenamefont{Slusky et~al.}(2001)\citenamefont{Slusky, Rogado,
  Regan, Hayward, Khalifah, He, Inumaru, Loureiro, Haas, Zandbergen
  et~al.}}]{cava1}
\bibinfo{author}{\bibfnamefont{J.~S.} \bibnamefont{Slusky}},
  \bibinfo{author}{\bibfnamefont{N.}~\bibnamefont{Rogado}},
  \bibinfo{author}{\bibfnamefont{K.~S.} \bibnamefont{Regan}},
  \bibinfo{author}{\bibfnamefont{M.~A.} \bibnamefont{Hayward}},
  \bibinfo{author}{\bibfnamefont{P.}~\bibnamefont{Khalifah}},
  \bibinfo{author}{\bibfnamefont{T.}~\bibnamefont{He}},
  \bibinfo{author}{\bibfnamefont{K.}~\bibnamefont{Inumaru}},
  \bibinfo{author}{\bibfnamefont{S.~M.} \bibnamefont{Loureiro}},
  \bibinfo{author}{\bibfnamefont{M.~K.} \bibnamefont{Haas}},
  \bibinfo{author}{\bibfnamefont{H.~W.} \bibnamefont{Zandbergen}},
  \bibnamefont{et~al.}, \bibinfo{journal}{Nature}
  \textbf{\bibinfo{volume}{410}}, \bibinfo{pages}{343} (\bibinfo{year}{2001}).

\bibitem[{\citenamefont{Profeta et~al.}(2003)\citenamefont{Profeta, Continenza,
  and Massidda}}]{profeta}
\bibinfo{author}{\bibfnamefont{G.}~\bibnamefont{Profeta}},
  \bibinfo{author}{\bibfnamefont{A.}~\bibnamefont{Continenza}},
  \bibnamefont{and} \bibinfo{author}{\bibfnamefont{S.}~\bibnamefont{Massidda}},
  \bibinfo{journal}{Phys. Rev. B} \textbf{\bibinfo{volume}{68}},
  \bibinfo{pages}{144508} (\bibinfo{year}{2003}).

\bibitem[{\citenamefont{Barabash and Stroud}(2002)}]{barabash}
\bibinfo{author}{\bibfnamefont{V.}~\bibnamefont{Barabash}} \bibnamefont{and}
  \bibinfo{author}{\bibfnamefont{D.}~\bibnamefont{Stroud}},
  \bibinfo{journal}{Phys. Rev. B} \textbf{\bibinfo{volume}{66}},
  \bibinfo{pages}{012509} (\bibinfo{year}{2002}).

\bibitem[{\citenamefont{Burkhardt et~al.}(2004)\citenamefont{Burkhardt, Gurin,
  Haarmann, H.~Borrmann, Yareskko, and Grin}}]{grin}
\bibinfo{author}{\bibfnamefont{U.}~\bibnamefont{Burkhardt}},
  \bibinfo{author}{\bibfnamefont{V.}~\bibnamefont{Gurin}},
  \bibinfo{author}{\bibfnamefont{F.}~\bibnamefont{Haarmann}},
  \bibinfo{author}{\bibfnamefont{W.~S.} \bibnamefont{H.~Borrmann}},
  \bibinfo{author}{\bibfnamefont{A.}~\bibnamefont{Yareskko}}, \bibnamefont{and}
  \bibinfo{author}{\bibfnamefont{Y.}~\bibnamefont{Grin}}, \bibinfo{journal}{J.
  Solid State Chem.} \textbf{\bibinfo{volume}{177}}, \bibinfo{pages}{389}
  (\bibinfo{year}{2004}).

\bibitem[{\citenamefont{Maurin et~al.}(2002)\citenamefont{Maurin, Margadonna,
  Prassides, Takenobu, Iwasa, and Fitch}}]{maurin}
\bibinfo{author}{\bibfnamefont{I.}~\bibnamefont{Maurin}},
  \bibinfo{author}{\bibfnamefont{S.}~\bibnamefont{Margadonna}},
  \bibinfo{author}{\bibfnamefont{K.}~\bibnamefont{Prassides}},
  \bibinfo{author}{\bibfnamefont{T.}~\bibnamefont{Takenobu}},
  \bibinfo{author}{\bibfnamefont{Y.}~\bibnamefont{Iwasa}}, \bibnamefont{and}
  \bibinfo{author}{\bibfnamefont{A.~N.} \bibnamefont{Fitch}},
  \bibinfo{journal}{Chem. Mater.} \textbf{\bibinfo{volume}{14}},
  \bibinfo{pages}{3894} (\bibinfo{year}{2002}).

\bibitem[{\citenamefont{Bharathi et~al.}(2002)\citenamefont{Bharathi,
  Balaselvi, Kalavathi, Reddy, Sastry, Hariharan, and
  Radhakrishnan}}]{bharathi}
\bibinfo{author}{\bibfnamefont{A.}~\bibnamefont{Bharathi}},
  \bibinfo{author}{\bibfnamefont{S.~J.} \bibnamefont{Balaselvi}},
  \bibinfo{author}{\bibfnamefont{S.}~\bibnamefont{Kalavathi}},
  \bibinfo{author}{\bibfnamefont{G.~L.~N.} \bibnamefont{Reddy}},
  \bibinfo{author}{\bibfnamefont{V.~S.} \bibnamefont{Sastry}},
  \bibinfo{author}{\bibfnamefont{Y.}~\bibnamefont{Hariharan}},
  \bibnamefont{and} \bibinfo{author}{\bibfnamefont{T.~S.}
  \bibnamefont{Radhakrishnan}}, \bibinfo{journal}{Physica C}
  \textbf{\bibinfo{volume}{370}}, \bibinfo{pages}{211} (\bibinfo{year}{2002}).

\bibitem[{\citenamefont{Schmidt et~al.}(2003)\citenamefont{Schmidt, Gray,
  Hinks, Zsadzinski, Avdeev, Jorgensen, and Burley}}]{schmidt}
\bibinfo{author}{\bibfnamefont{H.}~\bibnamefont{Schmidt}},
  \bibinfo{author}{\bibfnamefont{K.~E.} \bibnamefont{Gray}},
  \bibinfo{author}{\bibfnamefont{D.~G.} \bibnamefont{Hinks}},
  \bibinfo{author}{\bibfnamefont{J.~F.} \bibnamefont{Zsadzinski}},
  \bibinfo{author}{\bibfnamefont{M.}~\bibnamefont{Avdeev}},
  \bibinfo{author}{\bibfnamefont{J.~D.} \bibnamefont{Jorgensen}},
  \bibnamefont{and} \bibinfo{author}{\bibfnamefont{J.~C.}
  \bibnamefont{Burley}}, \bibinfo{journal}{Phys. Rev. B}
  \textbf{\bibinfo{volume}{060508}}, \bibinfo{pages}{68}
  (\bibinfo{year}{2003}).

\bibitem[{\citenamefont{Andeev et~al.}(2003)\citenamefont{Andeev, Jorgensen,
  Ribiero, Bud'ko, and Canfield}}]{adveev}
\bibinfo{author}{\bibfnamefont{M.}~\bibnamefont{Andeev}},
  \bibinfo{author}{\bibfnamefont{J.~D.} \bibnamefont{Jorgensen}},
  \bibinfo{author}{\bibfnamefont{R.~A.} \bibnamefont{Ribiero}},
  \bibinfo{author}{\bibfnamefont{S.~L.} \bibnamefont{Bud'ko}},
  \bibnamefont{and} \bibinfo{author}{\bibfnamefont{P.~C.}
  \bibnamefont{Canfield}}, \bibinfo{journal}{Physica C}
  \textbf{\bibinfo{volume}{301}}, \bibinfo{pages}{387} (\bibinfo{year}{2003}).

\bibitem[{\citenamefont{Ribiero et~al.}(2003)\citenamefont{Ribiero, Bud'ko,
  Petrovic, and Canfield}}]{ribiero}
\bibinfo{author}{\bibfnamefont{R.~A.} \bibnamefont{Ribiero}},
  \bibinfo{author}{\bibfnamefont{S.~L.} \bibnamefont{Bud'ko}},
  \bibinfo{author}{\bibfnamefont{C.}~\bibnamefont{Petrovic}}, \bibnamefont{and}
  \bibinfo{author}{\bibfnamefont{P.~C.} \bibnamefont{Canfield}},
  \bibinfo{journal}{Physica C} \textbf{\bibinfo{volume}{16}},
  \bibinfo{pages}{385} (\bibinfo{year}{2003}).

\bibitem[{\citenamefont{Masui et~al.}(2004)\citenamefont{Masui, Lee, and
  Tajima}}]{masui}
\bibinfo{author}{\bibfnamefont{T.}~\bibnamefont{Masui}},
  \bibinfo{author}{\bibfnamefont{S.}~\bibnamefont{Lee}}, \bibnamefont{and}
  \bibinfo{author}{\bibfnamefont{S.}~\bibnamefont{Tajima}},
  \bibinfo{journal}{cond-mat/0312458}  (\bibinfo{year}{2004}).

\bibitem[{\citenamefont{Wilke et~al.}(2004)\citenamefont{Wilke, Bud'ko,
  Canfield, Finnemore, Suplinskas, and Hannahs}}]{wilke}
\bibinfo{author}{\bibfnamefont{R.~H.~T.} \bibnamefont{Wilke}},
  \bibinfo{author}{\bibfnamefont{S.~L.} \bibnamefont{Bud'ko}},
  \bibinfo{author}{\bibfnamefont{P.~C.} \bibnamefont{Canfield}},
  \bibinfo{author}{\bibfnamefont{D.~K.} \bibnamefont{Finnemore}},
  \bibinfo{author}{\bibfnamefont{R.~J.} \bibnamefont{Suplinskas}},
  \bibnamefont{and} \bibinfo{author}{\bibfnamefont{S.~T.}
  \bibnamefont{Hannahs}}, \bibinfo{journal}{cond-mat/0312235}
  (\bibinfo{year}{2004}).

\bibitem[{\citenamefont{An and Pickett}(2001)}]{jan}
\bibinfo{author}{\bibfnamefont{J.~M.} \bibnamefont{An}} \bibnamefont{and}
  \bibinfo{author}{\bibfnamefont{W.~E.} \bibnamefont{Pickett}},
  \bibinfo{journal}{Phys. Rev. Lett.} \textbf{\bibinfo{volume}{86}},
  \bibinfo{pages}{4366} (\bibinfo{year}{2001}).

\bibitem[{\citenamefont{Holanova et~al.}(2004)\citenamefont{Holanova, Szabo,
  Samuely, Wilke, Bud'ko, and Canfield}}]{holanova}
\bibinfo{author}{\bibfnamefont{Z.}~\bibnamefont{Holanova}},
  \bibinfo{author}{\bibfnamefont{P.}~\bibnamefont{Szabo}},
  \bibinfo{author}{\bibfnamefont{P.}~\bibnamefont{Samuely}},
  \bibinfo{author}{\bibfnamefont{R.~H.~T.} \bibnamefont{Wilke}},
  \bibinfo{author}{\bibfnamefont{S.~L.} \bibnamefont{Bud'ko}},
  \bibnamefont{and} \bibinfo{author}{\bibfnamefont{P.~C.}
  \bibnamefont{Canfield}}, \bibinfo{journal}{cond-mat/0404096}
  (\bibinfo{year}{2004}).

\bibitem[{\citenamefont{Samuely et~al.}(2003)\citenamefont{Samuely, Holanova,
  Szabo, Kacmarcik, Ribiero, Bud'ko, and Canfield}}]{samuely}
\bibinfo{author}{\bibfnamefont{P.}~\bibnamefont{Samuely}},
  \bibinfo{author}{\bibfnamefont{Z.}~\bibnamefont{Holanova}},
  \bibinfo{author}{\bibfnamefont{P.}~\bibnamefont{Szabo}},
  \bibinfo{author}{\bibfnamefont{J.}~\bibnamefont{Kacmarcik}},
  \bibinfo{author}{\bibfnamefont{R.~A.} \bibnamefont{Ribiero}},
  \bibinfo{author}{\bibfnamefont{S.~L.} \bibnamefont{Bud'ko}},
  \bibnamefont{and} \bibinfo{author}{\bibfnamefont{P.~C.}
  \bibnamefont{Canfield}}, \bibinfo{journal}{Phys. Rev. B}
  \textbf{\bibinfo{volume}{68}}, \bibinfo{pages}{020505}
  (\bibinfo{year}{2003}).

\bibitem[{\citenamefont{Puzniak et~al.}(2004)\citenamefont{Puzniak, Angst,
  Szewczyk, Jun, Kazakov, and Karpinski}}]{puzniak}
\bibinfo{author}{\bibfnamefont{R.}~\bibnamefont{Puzniak}},
  \bibinfo{author}{\bibfnamefont{M.}~\bibnamefont{Angst}},
  \bibinfo{author}{\bibfnamefont{A.}~\bibnamefont{Szewczyk}},
  \bibinfo{author}{\bibfnamefont{J.}~\bibnamefont{Jun}},
  \bibinfo{author}{\bibfnamefont{S.~M.} \bibnamefont{Kazakov}},
  \bibnamefont{and}
  \bibinfo{author}{\bibfnamefont{J.}~\bibnamefont{Karpinski}},
  \bibinfo{journal}{cond-mat/0404579}  (\bibinfo{year}{2004}).

\bibitem[{\citenamefont{Avdeev et~al.}(2003)\citenamefont{Avdeev, Jorgensen,
  Robeiro, Bud'ko, and Canfield}}]{avdeev}
\bibinfo{author}{\bibfnamefont{M.}~\bibnamefont{Avdeev}},
  \bibinfo{author}{\bibfnamefont{J.~D.} \bibnamefont{Jorgensen}},
  \bibinfo{author}{\bibfnamefont{R.~A.} \bibnamefont{Robeiro}},
  \bibinfo{author}{\bibfnamefont{S.~L.} \bibnamefont{Bud'ko}},
  \bibnamefont{and} \bibinfo{author}{\bibfnamefont{P.~C.}
  \bibnamefont{Canfield}}, \bibinfo{journal}{Physica C}
  \textbf{\bibinfo{volume}{387}}, \bibinfo{pages}{301} (\bibinfo{year}{2003}).

\bibitem[{\citenamefont{Yan and Al-Jassim}(2002)}]{yanfa}
\bibinfo{author}{\bibfnamefont{Y.}~\bibnamefont{Yan}} \bibnamefont{and}
  \bibinfo{author}{\bibfnamefont{M.~M.} \bibnamefont{Al-Jassim}},
  \bibinfo{journal}{J. Appl. Phys.} \textbf{\bibinfo{volume}{92}},
  \bibinfo{pages}{7687} (\bibinfo{year}{2002}).

\bibitem[{\citenamefont{Erwin and Mazin}(2003)}]{erwin}
\bibinfo{author}{\bibfnamefont{S.~C.} \bibnamefont{Erwin}} \bibnamefont{and}
  \bibinfo{author}{\bibfnamefont{I.~I.} \bibnamefont{Mazin}},
  \bibinfo{journal}{Physical Review B} \textbf{\bibinfo{volume}{68}},
  \bibinfo{pages}{132505} (\bibinfo{year}{2003}).

\bibitem[{\citenamefont{Singh}(2003)}]{ppsingh}
\bibinfo{author}{\bibfnamefont{P.~P.} \bibnamefont{Singh}},
  \bibinfo{journal}{cond-mat/0304436}  (\bibinfo{year}{2003}).

\bibitem[{\citenamefont{An et~al.}(2003)\citenamefont{An, Rosner, Savrasov, and
  Pickett}}]{libc1}
\bibinfo{author}{\bibfnamefont{J.~M.} \bibnamefont{An}},
  \bibinfo{author}{\bibfnamefont{H.}~\bibnamefont{Rosner}},
  \bibinfo{author}{\bibfnamefont{S.~Y.} \bibnamefont{Savrasov}},
  \bibnamefont{and} \bibinfo{author}{\bibfnamefont{W.~E.}
  \bibnamefont{Pickett}}, \bibinfo{journal}{Physica B}
  \textbf{\bibinfo{volume}{328}}, \bibinfo{pages}{1} (\bibinfo{year}{2003}).

\bibitem[{\citenamefont{Pickett et~al.}(2003)\citenamefont{Pickett, An, Rosner,
  and Savrasov}}]{2D}
\bibinfo{author}{\bibfnamefont{W.~E.} \bibnamefont{Pickett}},
  \bibinfo{author}{\bibfnamefont{J.~M.} \bibnamefont{An}},
  \bibinfo{author}{\bibfnamefont{H.}~\bibnamefont{Rosner}}, \bibnamefont{and}
  \bibinfo{author}{\bibfnamefont{S.~Y.} \bibnamefont{Savrasov}},
  \bibinfo{journal}{Physica C} \textbf{\bibinfo{volume}{387}},
  \bibinfo{pages}{117} (\bibinfo{year}{2003}).

\bibitem[{\citenamefont{Kong et~al.}(2001)\citenamefont{Kong, Dolgov, Jepsen,
  and Andersen}}]{kong}
\bibinfo{author}{\bibfnamefont{Y.}~\bibnamefont{Kong}},
  \bibinfo{author}{\bibfnamefont{O.~V.} \bibnamefont{Dolgov}},
  \bibinfo{author}{\bibfnamefont{O.}~\bibnamefont{Jepsen}}, \bibnamefont{and}
  \bibinfo{author}{\bibfnamefont{O.~K.} \bibnamefont{Andersen}},
  \bibinfo{journal}{Phys. Rev. B} \textbf{\bibinfo{volume}{64}},
  \bibinfo{pages}{020501} (\bibinfo{year}{2001}).

\bibitem[{\citenamefont{Boeri et~al.}(2002)\citenamefont{Boeri, Bachelet,
  Cappelluti, and Pietronero}}]{boeri}
\bibinfo{author}{\bibfnamefont{L.}~\bibnamefont{Boeri}},
  \bibinfo{author}{\bibfnamefont{G.~B.} \bibnamefont{Bachelet}},
  \bibinfo{author}{\bibfnamefont{E.}~\bibnamefont{Cappelluti}},
  \bibnamefont{and}
  \bibinfo{author}{\bibfnamefont{L.}~\bibnamefont{Pietronero}},
  \bibinfo{journal}{Phys. Rev. B} \textbf{\bibinfo{volume}{65}},
  \bibinfo{pages}{214501} (\bibinfo{year}{2002}).

\bibitem[{\citenamefont{Pickett}(2003)}]{wep}
\bibinfo{author}{\bibfnamefont{W.~E.} \bibnamefont{Pickett}},
  \bibinfo{journal}{Braz. J. Phys.} \textbf{\bibinfo{volume}{33}},
  \bibinfo{pages}{695} (\bibinfo{year}{2003}).

\bibitem[{\citenamefont{Blaha et~al.}(2002)\citenamefont{Blaha, Schwarz,
  Madsen, Kvasnicka, and Luitz}}]{wien}
\bibinfo{author}{\bibfnamefont{P.}~\bibnamefont{Blaha}},
  \bibinfo{author}{\bibfnamefont{K.}~\bibnamefont{Schwarz}},
  \bibinfo{author}{\bibfnamefont{G.~K.~H.} \bibnamefont{Madsen}},
  \bibinfo{author}{\bibfnamefont{D.}~\bibnamefont{Kvasnicka}},
  \bibnamefont{and} \bibinfo{author}{\bibfnamefont{J.}~\bibnamefont{Luitz}},
  \bibinfo{journal}{J. Phys. Chem. Sol.} \textbf{\bibinfo{volume}{63}},
  \bibinfo{pages}{2201} (\bibinfo{year}{2002}).

\bibitem[{\citenamefont{Perdew et~al.}(1996)\citenamefont{Perdew, Burke, and
  Ernzerhof}}]{gga}
\bibinfo{author}{\bibfnamefont{J.~P.} \bibnamefont{Perdew}},
  \bibinfo{author}{\bibfnamefont{K.}~\bibnamefont{Burke}}, \bibnamefont{and}
  \bibinfo{author}{\bibfnamefont{M.}~\bibnamefont{Ernzerhof}},
  \bibinfo{journal}{PRL} \textbf{\bibinfo{volume}{77}}, \bibinfo{pages}{3865}
  (\bibinfo{year}{1996}).

\bibitem[{\citenamefont{Sjostedt et~al.}(2000)\citenamefont{Sjostedt,
  Nordstrom, and Singh}}]{apw}
\bibinfo{author}{\bibfnamefont{E.}~\bibnamefont{Sjostedt}},
  \bibinfo{author}{\bibfnamefont{L.}~\bibnamefont{Nordstrom}},
  \bibnamefont{and} \bibinfo{author}{\bibfnamefont{D.~J.} \bibnamefont{Singh}},
  \bibinfo{journal}{Solid State Commun.} \textbf{\bibinfo{volume}{114}},
  \bibinfo{pages}{15} (\bibinfo{year}{2000}).

\bibitem[{\citenamefont{Koepernik and Eschrig}(1999)}]{klaus1}
\bibinfo{author}{\bibfnamefont{K.}~\bibnamefont{Koepernik}} \bibnamefont{and}
  \bibinfo{author}{\bibfnamefont{H.}~\bibnamefont{Eschrig}},
  \bibinfo{journal}{Phys. Rev. B} \textbf{\bibinfo{volume}{59}},
  \bibinfo{pages}{1743} (\bibinfo{year}{1999}).

\bibitem[{\citenamefont{Koepernik et~al.}(1997)\citenamefont{Koepernik,
  Velicky, Hayn, and Eschrig}}]{klaus2}
\bibinfo{author}{\bibfnamefont{K.}~\bibnamefont{Koepernik}},
  \bibinfo{author}{\bibfnamefont{B.}~\bibnamefont{Velicky}},
  \bibinfo{author}{\bibfnamefont{R.}~\bibnamefont{Hayn}}, \bibnamefont{and}
  \bibinfo{author}{\bibfnamefont{H.}~\bibnamefont{Eschrig}},
  \bibinfo{journal}{Phys. Rev. B} \textbf{\bibinfo{volume}{55}},
  \bibinfo{pages}{5717} (\bibinfo{year}{1997}).

\bibitem[{\citenamefont{Blackman et~al.}(1971)\citenamefont{Blackman,
  Esterling, and Berk}}]{BEB}
\bibinfo{author}{\bibfnamefont{J.~A.} \bibnamefont{Blackman}},
  \bibinfo{author}{\bibfnamefont{D.~M.} \bibnamefont{Esterling}},
  \bibnamefont{and} \bibinfo{author}{\bibfnamefont{N.~F.} \bibnamefont{Berk}},
  \bibinfo{journal}{Phys. Rev. B} \textbf{\bibinfo{volume}{4}},
  \bibinfo{pages}{2412} (\bibinfo{year}{1971}).

\bibitem[{\citenamefont{Agrestini et~al.}()\citenamefont{Agrestini, Metallo,
  Filippi, Simonelli, Campi, Sanipoli, Liarokapis, Negri, Giovannini, Saccone
  et~al.}}]{bianconi}
\bibinfo{author}{\bibfnamefont{S.}~\bibnamefont{Agrestini}},
  \bibinfo{author}{\bibfnamefont{C.}~\bibnamefont{Metallo}},
  \bibinfo{author}{\bibfnamefont{M.}~\bibnamefont{Filippi}},
  \bibinfo{author}{\bibfnamefont{L.}~\bibnamefont{Simonelli}},
  \bibinfo{author}{\bibfnamefont{G.}~\bibnamefont{Campi}},
  \bibinfo{author}{\bibfnamefont{C.}~\bibnamefont{Sanipoli}},
  \bibinfo{author}{\bibfnamefont{E.}~\bibnamefont{Liarokapis}},
  \bibinfo{author}{\bibfnamefont{S.~D.} \bibnamefont{Negri}},
  \bibinfo{author}{\bibfnamefont{M.}~\bibnamefont{Giovannini}},
  \bibinfo{author}{\bibfnamefont{A.}~\bibnamefont{Saccone}},
  \bibnamefont{et~al.}, \bibinfo{journal}{cond-mat/0408095}. 

\end{thebibliography}

\end{document}